\documentclass[twocolumn]{aa} 

\usepackage{graphicx}
\usepackage{txfonts}
\usepackage{natbib}
\usepackage{rotating}
\usepackage{color}
\usepackage{longtable,lscape}
\newcommand{\xspec}{{\it XSPEC}}

\begin{document}

\title{New XMM-Newton observation of the Phoenix cluster: properties of the cool core}



\author{P. Tozzi\inst{1}, F. Gastaldello\inst{2}, S. Molendi\inst{2},
S. Ettori\inst{3,4}, J. S. Santos\inst{1}, S. De Grandi\inst{5},  I. Balestra\inst{6}, 
P. Rosati\inst{7,1}, B. Altieri\inst{8}, G. Cresci\inst{1}, F. Menanteau\inst{9,10}, I. Valtchanov\inst{8}}


\institute{
INAF, Osservatorio Astrofisico di Firenze, Largo Enrico Fermi 5, I-50125, Firenze, Italy\\
e-mail: ptozzi@arcetri.astro.it
\and INAF, IASF Milano, via E. Bassini 15 I-20133 Milano, Italy 
\and INAF, Osservatorio Astronomico di Bologna, viale Berti Pichat 6/2, I–40127 Bologna, Italy 
\and INFN, Sezione di Bologna, viale Berti Pichat 6/2, I-40127 Bologna, Italy
\and INAF, Osservatorio Astronomico di Brera, via E. Bianchi 46, I-23807 Merate, Italy 
\and  INAF, Osservatorio Astronomico di Trieste, via G.B. Tiepolo 11, I–34131, Trieste, Italy 
\and Universit\`a degli Studi di Ferrara, Via Savonarola, 9 - 44121 Ferrara , Italy 
\and European Space Astronomy Centre (ESAC), European Space Agency, Apartado de Correos 78, 28691 Villanueva de la Ca\~nada, Madrid, Spain 
\and National Center for Supercomputing Applications, University of Illinois at Urbana-Champaign, 1205 W. Clark St., Urbana, IL 61801, USA 
\and  Department of Astronomy, University of Illinois at Urbana-Champaign, W. Green Street, Urbana, IL 61801, USA 
}

\date{Received 26 January 2015 /  Accepted 25 May 2015}

\abstract
{}
{We present a spectral analysis of a deep (220 ks) XMM-Newton observation of the 
Phoenix cluster (SPT-CL J2344-4243).  We also use  {{\sl {Chandra}}} archival ACIS-I 
data that are useful for modeling the properties
of the central bright active galactic nucleus and global intracluster medium.} 
{We extracted CCD and reflection grating spectrometer (RGS) X-ray spectra from the core region to search for the signature of 
cold gas and to finally constrain the mass deposition rate in the cooling flow  that  
is thought to be responsible for the massive star formation episode observed in the brightest cluster galaxy (BCG). }
{We find an average mass-deposition rate of $\dot M = 620 \, (-190 \, +200)_{stat} \,\, (-50 \, +150)_{syst}
M_\odot$ yr$^{-1}$ in the temperature range 0.3-3.0 keV from MOS data.  
A temperature-resolved analysis shows that a significant amount of gas is deposited at about 1.8 keV and above, while only 
upper limits on the order of hundreds of $M_\odot$ yr$^{-1}$ can be placed in the 0.3-1.8
keV temperature range.   From pn data we obtain $\dot M = 210 \, (-80 \, 
+85)_{stat} \,\, ( -35\,\, +60)_{syst} M_\odot$ yr$^{-1}$ in the  $0.3-3.0$ keV  temperature range, 
while the upper limits from the temperature-resolved analysis are typically a factor of 3 lower than MOS data.
No line emission from ionization states below Fe XXIII is seen above $12 \AA  $ in the RGS spectrum, and
the amount of gas cooling below $\sim 3$ keV has a formal best-fit value
$\dot M = 122_{-122}^{+343}$ $M_{\odot}$ yr$^{-1}$.  In addition, our analysis of the far-infrared  spectral energy
distribution of the BCG based on  Herschel data provides a star formation rate (SFR)  equal to $530 \, M_\odot$ yr$^{-1}$ 
with an uncertainty of 10\%, which is lower than previous estimates by a factor 1.5.
Overall, current limits on the mass deposition rate from MOS data are consistent with  the SFR
observed in the BCG, while pn data prefer a lower value of $\dot M \sim SFR/3$, which is inconsistent with the SFR at the $3\sigma$ confidence level. 
}
{Current data are able to firmly identify a substantial amount of cooling gas only above 1.8 keV in the core of
the Phoenix cluster.  At lower temperatures, the upper limits on $\dot M$ from MOS and pn data differ by a factor of 3.  While 
the MOS data analysis is consistent with values as high as $\dot M \sim 1000$ within $1 \, \sigma$, pn data provide 
$\dot M < 500 M_\odot$ yr$^{-1}$  at $3\sigma$ confidence level at a temperature below 1.8 keV.  At present, this discrepancy cannot be explained on the basis
of known calibration uncertainties  or other sources of statistical noise.
}


\titlerunning{XMM--Newton observation of the Phoenix cluster}

\authorrunning{Tozzi et al.}

\maketitle

\section{Introduction}

The majority of baryons in clusters of galaxies is constituted by  
virialized hot gas \citep{2003Lin,2013Gonzalez} that emits X-ray 
via thermal bremsstrahlung.  Temperature, density, and chemical
composition of the so-called intracluster medium (ICM)  can be directly
measured thanks to X-ray imaging and spectroscopic observations.  
Spatially resolved spectroscopic studies showed a significant temperature decrease and strongly peaked surface brightness profiles
in the center of a significant fraction of the cluster population.  The short cooling
times associated with these high-density gas regions led to the 
conclusion that a massive cooling flow was developing in the ICM in most
of the clusters \citep{1976Silk,1977Cowie,1977Fabian,1978Mathews}.  The fate of this cooling gas 
would be to feed massive star formation episodes.

On the basis of the isobaric cooling-flow model \citep{1977Fabian,1994Fabian}, it was estimated that typical cooling flows may 
develop mass deposition rates in the range of a few $\times 100 - 1000 \, M_{\sun}$ yr$^{-1}$.
However, the lack of massive star formation events and of large reservoirs of cold gas 
at the center of galaxy clusters casts some doubts on the hypothesis of a complete cooling of the ICM in cluster cores.  
The picture changed dramatically when X-ray observations, in particular of the RGS
instrument onboard the XMM-Newton telescope, revealed a severe
deficit of emission lines compared to the predictions of the isobaric
cooling--flow model in all the groups and clusters of galaxies
with putative cooling flows \citep{2001Tamura, 2001Peterson,2001Kaastra,
2003Peterson}.  Interestingly, this result has also been found in XMM-Newton and {\sl Chandra} CCD spectra, 
despite the lower resolution, thanks to the prominent complex of iron L-shell lines
\citep{2000McNamara,2001Boehringer,2001Molendi,2002Boehringer}. This directly implies that the cooling gas is
present only in small amounts, typically ten times lower than expected
for steady-state, isobaric radiative cooling \citep[see][]{2006Peterson}.  
This determined a change from the cooling-flow paradigm, with typical deposition rates
of about $100-1000 M_\odot$ yr$^{-1}$, to the cool-core paradigm, where most of the gas
is kept at temperatures higher than one-third of the ambient cluster temperature, and the mass deposition
rate, if any, is due to a residual cooling flow of about few tens  $M_\odot$ yr$^{-1}$.

Another direct implication of these observations is that there must be some process that heats
the gas and  prevents its cooling.  Among the many mechanisms investigated in the past years, 
feedback from active galactic nuclei (AGN) is considered the most plausible heating source.   Radio AGN are ubiquitous in cool cores \citep[see][]{2009Sun},
and interactions between the radio jets and the ICM have been observed unambiguously.   AGN outbursts can in principle inject sufficient energy into 
the ICM \citep[see][]{2005McNamara}.  The relativistic electrons in jets associated with the
central cluster galaxy are able to carve large cavities into the ICM.  The free energy associated
with these bubbles is plausibly transferred into the ICM and thermalized through turbulence
\citep[see][]{2012McNamara,2014Zhuravleva}.  In addition, there is increasing evidence of interactions of
AGN outflows  with metal-rich gas along the cavities and edges of radio jets of some individual clusters and groups \citep{2011Kirkpatrick,2013Ettori_CF}, consistent
with numerical simulations showing that AGN outflows are able to advect ambient, iron-rich material from the core to a few hundred kpc
away \citep[e.g.,][]{2011Gaspari_a,2011Gaspari_b}.   The feedback mechanism has been observed in its full complexity in nearby clusters such as
Perseus \citep{2003aFabian,2006Fabian,2011Fabian}, Hydra A \citep{2000McNamara}, 
and a few other clusters \citep[see][]{2011Blanton}. 

In addition, the regular behavior of cool cores is not only observed in local clusters, but seems to hold up to high redshifts.  High angular resolution
observations of cool cores at $z\leq 1$ performed by our group \citep{2010Santos,2012Santos}
showed that the radio feedback mechanism is already present  at $z\sim 1$.  Temperature and metallicity
profiles in cool cores are broadly consistent with local ones, with a remarkable difference in the 
the metal distribution that appears to be more concentrated in the core than in local clusters  \citep{2014DeGrandi}.  This indicates
that radio feedback also plays a role in the spatial distribution of metals.  The regularity of the cool-core appearance over the entire cluster
population and a wide range of epochs points toward a gentle heating mechanism, with the radio AGN acting with a short duty-cycle
to counterbalance the onset of cooling flows since the very first stages of cluster formation.
At the same time, outburst shocks may provide a more violent heating mechanism.
It is poorly understood whether AGN heating of the ICM occurs violently through
shocks or through bubbles in pressure equilibrium that cause turbulence. 
Outburst shock with jumps in temperature are very hard to detect, and shocks
have been unambiguously detected in only a few cases \citep[see, e.g., A2052 and NGC5813,][]{2011Blanton,2011Randall}.
It is now widely accepted, however, that the mechanical energy provided by the AGN through jets is
sufficient to overcome the cooling process in cluster cores.  Therefore questions remain  about the detailed physical mechanism that transfers energy
to the ICM, and how this mechanism gives rise to the regular cool-core thermal structure, which requires a minimum temperature
a factor $\sim 3$ lower than the ambient cluster temperature.

Surprisingly, a recent observation introduced further changes in the picture
outlined here.  The SZ-selected cluster SPT-CLJ2344-4243 \citep[also known as the Phoenix cluster,][]{2012McDonald} at $z\sim 0.596$
for the first time shows hints of a massive cooling-flow-induced starburst, 
suggesting that the feedback source responsible for preventing runaway 
cooling may not yet be fully established.  SPT--CLJ2344-4243 shows a strong cool core with a
potential mass deposition rate of $\sim 3000 M_\odot$ yr$^{-1}$ derived from
the X-ray luminosity.  \citet{2012McDonald} argued that the 
Phoenix cluster might harbor an almost isobaric cooling flow with an unusually high mass deposition rate. 
The strongest hint comes from the very high star formation rate (SFR) observed in the brightest cluster galaxy (BCG), which has originally been estimated to be $\sim 700 M_\odot$ yr$^{-1}$, with large  
$1 \sigma$ errors ranging from $200$ to $500 M_\odot$ yr$^{-1}$ \citep[][]{2012McDonald}.  
However, an accurate measurement of the SFR is made difficult by the 
presence of a  strongly absorbed AGN, whose contribution to the BCG emission
can be accounted for in different ways.  Recently, the HST/WFC3 observation of the Phoenix 
\citep{2013aMcDonald} showed filamentary blue emission out to 40 kpc and beyond, and the estimated, 
extinction-corrected SFR has been updated to a more accurate value of 
$798\pm 42 M_{\odot}$ yr$^{-1}$, consistent with optical and IR data at lower spatial resolution.

In this work we present the analysis of a 220 ks observation with XMM-Newton 
awarded in AO12 on the Phoenix cluster, with the main goal  of 
investigating the thermal structure of the cool core  and comparing the mass deposition rate  in the core
to the star formation rate in the BCG.  We also use archival {\it Chandra} data (about 10 ks with ACIS-I) to model the emission of the AGN in the BCG
and global ICM properties.  Finally, we revise the SFR in the BCG on the basis of far-IR (FIR) data from the   {\sl Herschel} Observatory.

The paper is organized as follows.  In Sect. \ref{reduction} we describe the reduction of the
XMM-Newton and {\sl Chandra} data.  In Sect. \ref{spectral_Chandra} we present the results from the
{\sl Chandra} data analysis on the central AGN spectrum and the global ICM properties.
In Sect. \ref{results} we describe our analysis strategy and present the results on the cool-core
temperature structure from EPIC MOS and pn data, RGS data, and {\sl Chandra} data.   
In Sect. \ref{rev_SFR} we revise the measurement of the star formation rate in the BCG from FIR data in view of a comparison 
with the mass deposition rate.  Finally, our conclusions are summarized
in Sect. \ref{conclusions}.  Throughout the paper, we adopt the seven-year WMAP cosmology
with $\Omega_{\Lambda} =0.73 $, $\Omega_m =0.27$, and $H_0 = 70.4 $ km
s$^{-1}$ Mpc$^{-1}$ \citep{2011Komatsu}.  Quoted
errors and upper limits always correspond to a 1 $\sigma$ confidence level, unless stated otherwise.

\begin{table*}
\centering
\caption{\label{exptime}XMM-Newton data: exposure times for each Obsid after data reduction.}
\begin{tabular}{|c|c|c|}
\hline
            &                                &                           \\
Obsid   & EPIC detector           &   effective $t_{exp}$     \\
            &                                &          ks                       \\
\hline
            &                                &                           \\
0722700101   &      MOS1      &              128.0                  \\
0722700101   &      MOS2      &              128.0                   \\
0722700101   &      pn           &              103.0                   \\
            &                                &                           \\
0722700201   &      MOS1      &            92.0                    \\
0722700201   &      MOS2      &             92.0                    \\
0722700201   &      pn           &              81.5                   \\
            &                                &                           \\
0693661801   &      MOS1      &            16.0                    \\
0693661801   &      MOS2      &             16.5                    \\
0693661801   &      pn           &              8.6                   \\
            &                                &                           \\
 \hline
\end{tabular}
\end{table*}

\section{Data reduction\label{reduction}}

\subsection{XMM-Newton: EPIC data}

We obtained a total of 225 ks with XMM-Newton on the Phoenix cluster in AO12
\footnote{Proposal ID 72270, {\sl "The thermal structure of the cool core in the Phoenix cluster"}, 
PI P. Tozzi}. Data were acquired in November and December 2013 (Obsid 0722700101, 132 ks,  
and 0722700201, 93 ks).  We added  a shorter 20 ks archival observation taken in 2012 to our analysis. 
\footnote{Proposal ID 069366, PI M. Arnaud.}
 
The observation data files (ODF) were processed to produce calibrated event files using the most recent release of the XMM-Newton Science Analysis
System (SAS v14.0.0),  with the calibration release XMM-CCF-REL-323, and running the tasks EPPROC and EMPROC for the pn and MOS, respectively, to generate calibrated and concatenated EPIC event lists.
Then, we filtered EPIC event lists for bad pixels, bad columns, cosmic-ray events outside the field of view (FOV), photons in the gaps (FLAG=0),
and applied standard grade selection,  corresponding to  PATTERN $<12$ for MOS and PATTERN $<=4$ for pn.
We removed soft proton flares by applying a threshold on the count rate in the $10$-$12$ keV energy band.
To define low-background intervals, we used the condition $RATE \leq 0.35$ for MOS and
$RATE \leq 0.4$ for pn.  We found  that for Obsid 0722700101, we have about 128 ks for MOS1 and MOS2 and 103 ks for pn after
data reduction.  For Obsid 0722700201 we have 92 ks for MOS1 and MOS2 and 81.5 ks for pn.
This means that removing high-background intervals reduces the effective  total time from 225 ks to 220 for MOS
(a loss of 2\% of the total time) and to 184.5 ks for pn (a loss of 18\% of the total time).
The archival data, Obsid 0693661801, were affected by flares by a larger amount.  From 20 ks of exposure
in Obsid 0693661801, we have 16 and 16.5 ks for MOS1 and MOS2, respectively (a loss of about 18-20\%),
and 8.6 ks for pn (a loss of about 60\% of the total time).

For each Obsid we merged the event files MOS1 and MOS2 to create
a single event file.  This procedure has been adopted because the MOS cameras typically yield mutually consistent fluxes over their whole energy
bandpass \footnote{See http://xmm2.esac.esa.int/docs/documents/CAL-TN-0018.pdf.}.  
Finally, we also removed out-of-time events from the pn event file and
spectra.   In Table 1, we list the resulting clean exposure times for the pn and the MOS
detectors for the three exposures used in this work.

Effective area and response matrix files were generated with the tasks  {\tt arfgen} and {\tt rmfgen}, respectively, for each obsid.
For MOS, effective area and response matrix files were computed for each detector and were eventually summed with a weight corresponding to the
effective exposure time of MOS1 and MOS2.

\subsection{XMM-Newton: RGS data}

We reduced the  RGS  data set using the standard SAS v14.0.0 pipeline processing through the RGSPROC tool. 
We filtered soft proton flares by excluding time periods where the count rate on CCD9\footnote{This is the CCD that generally records the fewest source events because of its 
location close to the optical axis and is the most susceptible to proton events.} is lower than 0.1 cts/s in a region free of source emission. 
The resulting effective exposure times, after removing flares from the data, are listed in Table \ref{tab:exp}.  As background we adopted the model background spectrum created
by the SAS task RGSBKGMODEL, which can be applied to a given observation 
from a combination of observations of empty fields, based on the count rate of the off-axis source-free region of 
CCD9.  We verified that the background spectrum obtained in this way is entirely consistent with a local background
extracted from beyond 98\% of the RGS point spread function (PSF). Finally, we focused on the first-order spectra and combined
spectra, backgrounds, and responses from all observations and from both RGS instruments using the SAS task 
RGSCOMBINE. 

\begin{table}[ht]
\caption{Cleaned exposure times for RGS detectors}
\label{tab:exp}
\centering
\begin{tabular}{|c|c|c|}
\hline
            &                                &                           \\
Obsid   & RGS detector           &   effective $t_{exp}$     \\
            &                                &          ks                       \\
\hline
            &                                &                           \\
0722700101 &  RGS1 & 126.4 \\
0722700101 &  RGS2 & 125.2 \\
            &                                &                           \\
0722700201 &  RGS1 & 92.4 \\
0722700201 &  RGS2 & 92.7 \\
            &                                &                           \\
\hline
\end{tabular}
\end{table}

\subsection{Archival {\sl Chandra} data}

The Phoenix cluster has been observed with ACIS-I for 11.9 ks in the VFAINT mode 
(Obsid 13401)\footnote{Proposal ID 13800933, {\sl "Chandra Observation of the Most Massive Galaxy 
Clusters Detected in the South Pole Telescope Survey"}, PI. G. Garmire.}.  
We performed a standard data reduction starting from the level=1 event
files, using the {\tt CIAO 4.6} software package, with the most recent
version of the {\sl Chandra} Calibration Database ({\tt CALDB 4.6.3}).  
We ran the task {\tt acis$\_$process$\_$events} to flag background events that are most
likely associated with cosmic rays and removed them.   
With this procedure, the ACIS particle background can be
significantly reduced compared to the standard grade selection.  The
data were filtered to include only the standard event grades 0, 2, 3, 4,
and 6.  We visually checked  for hot columns that were left after the standard
reduction.  As expected for exposures taken in VFAINT mode, we did not
find hot columns or flickering pixels after filtering out bad events.  Finally, we 
filtered time intervals with high background by performing a $3\sigma$
clipping of the background level using the script {\tt
analyze\_ltcrv}.   Only a negligible fraction of the exposure time was lost in this
step, and the final effective exposure time is 11.7 ks.  We remark that our spectral analysis 
will not be affected by possible undetected flares, since we are able to compute
the background in the same observation from a large  source-free region close to the cluster position, thus
taking into account any possible spectral distortion of the background
itself induced by the flares.

\section{Spectral analysis of {\sl Chandra} data: AGN spectrum and global ICM properties\label{spectral_Chandra}}

\begin{figure*}
\begin{center}
\includegraphics[width=6.5cm]{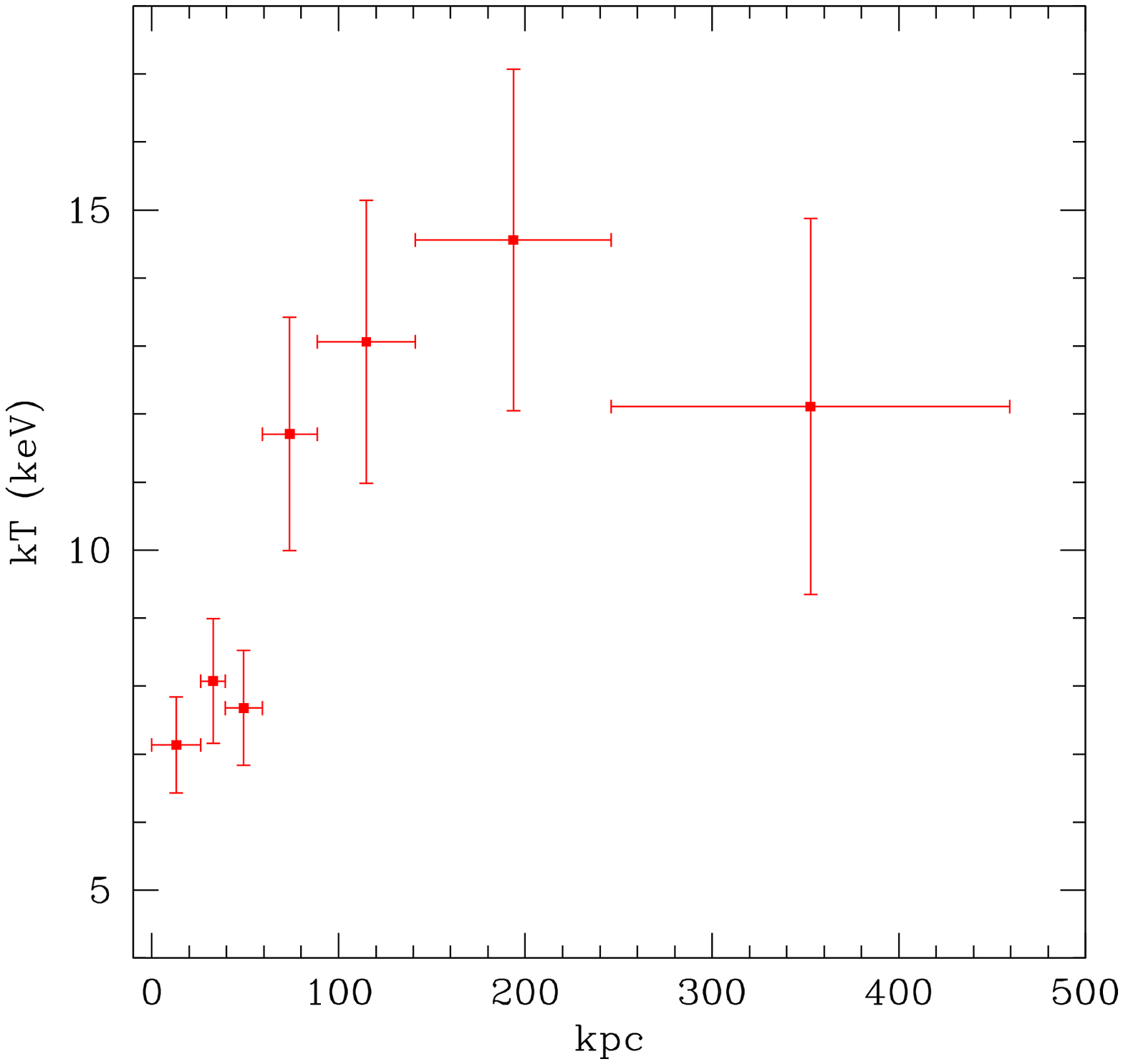}
\includegraphics[width=6.5cm]{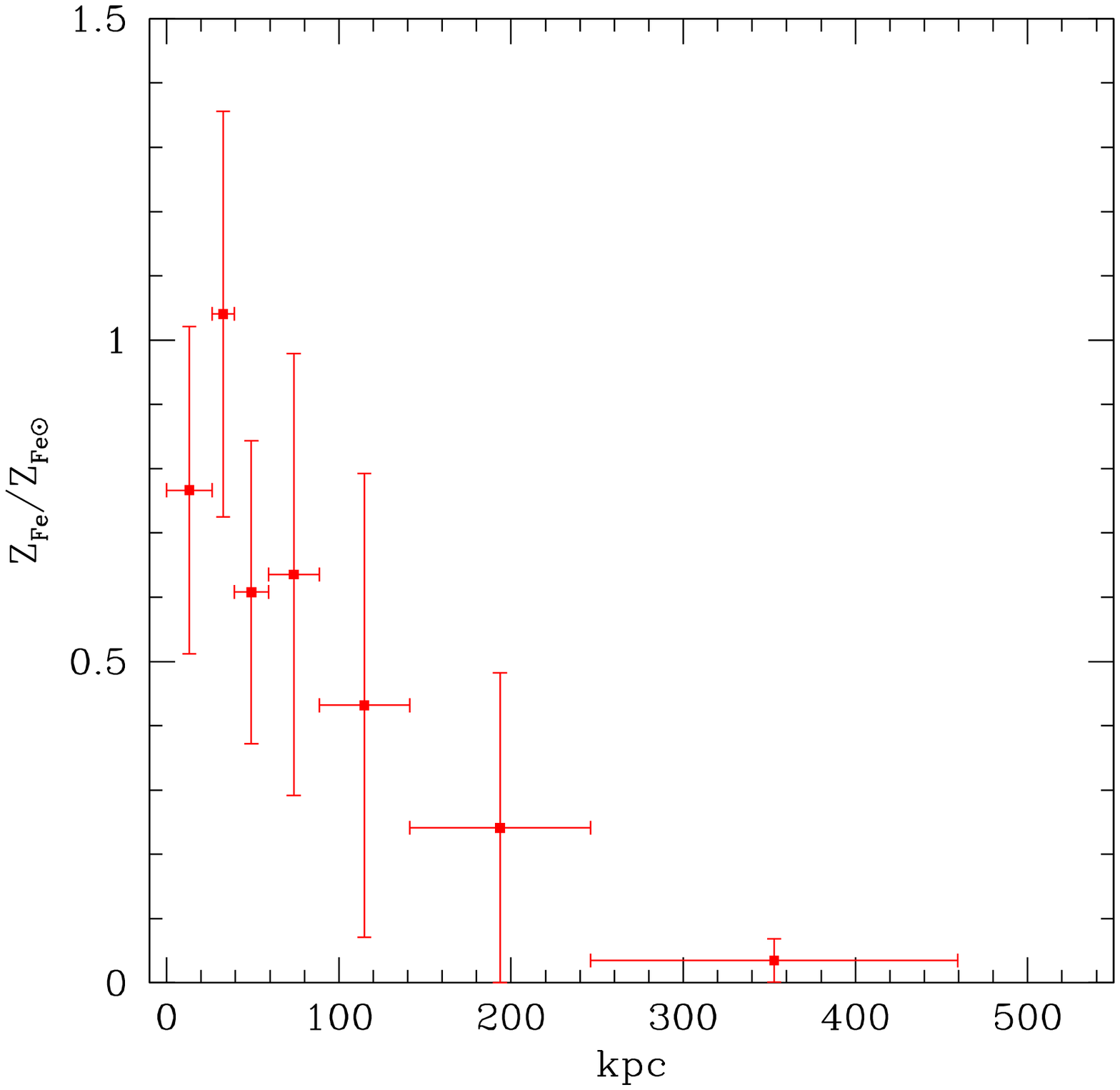}
\end{center}
\caption{
\label{chandraprofiles}
{\it Left panel}: projected temperature profile of SPT-CLJ2344 obtained from the 
11.7 ks {\sl Chandra} ACIS-I observation, after removing the central AGN. Errorbars
correspond to 1 $\sigma$.  
{\it Right panel}: projected iron abundance profile of SPT-CLJ2344 from {\sl Chandra} data.}
\end{figure*}

We first performed the spectral analysis of the {\sl Chandra} data.   Despite the short exposure time\footnote{A new 100 ks
observation with {\sl Chandra} has been taken in August 2014, PI M. McDonald.}, 
the high angular resolution of {\sl Chandra} can provide important parameters  useful in  the
spectral analysis of the XMM-Newton data.  In particular, it
is immediately clear from the high-resolution hard-band image that the BCG of the Phoenix cluster hosts a powerful obscured AGN.  In XMM-Newton data, 
the point spread function has an half-energy width (HEW) of about 15" at the aimpoint, which causes  
the AGN emission to be spread over the entire starburst region.  
{\sl Chandra} data allow us to accurately measure the spectrum of the central AGN and eventually
model its emission in the analysis of the XMM data.

We extracted a circular region with a radius of 1.5 arcsec to analyze the position of the central source
in the hard-band {\sl Chandra} image ($RA=23$:$44$:$43.9$, $DEC=-42$:$43$:$12.64$).  At variance with most cool core clusters, the 
AGN in the center of the Phoenix is extremely X-ray luminous.  We only considered the
energy range 1.0-10 keV for fitting purposes to avoid residual contamination from the 
thermal emission of the ICM.   We detected 575 net counts in the 1.0-7 keV band, most of them in the hard 2-7 keV band.
In our fit we fixed the redshift to the optical value $z_{opt} = 0.596$ \citep[see][]{2012McDonald}.
We also fixed the Galactic absorption to the value $NH_{Gal} = 1.52 \times 10^{20}$ cm$^{-2}$  
obtained from the radio map of  \citet{2005LAB} at the position of the cluster.  The AGN was modeled
as a power law with an intrinsic absorption (\xspec \, model {\tt zwabs $\times$ pow}) convolved
by the Galactic absorption model ({\tt tbabs}).  The slope of the power law was frozen to the 
value $\Gamma = 1.8$.

\begin{figure*}
\begin{center}
\includegraphics[angle=270,width=12cm]{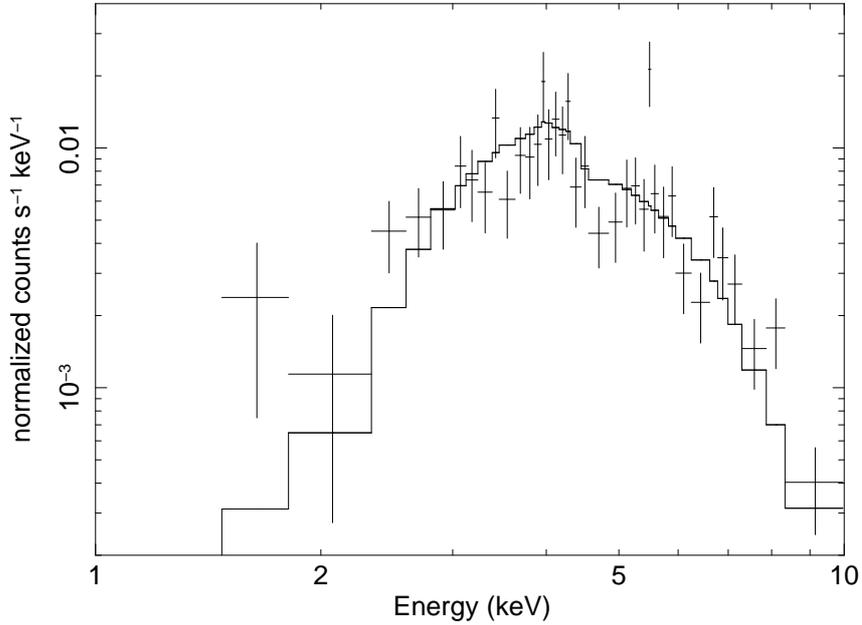}
\end{center}
\caption{
\label{AGNspectrum}Folded spectrum with best-fit model for the AGN in the center of the BCG of the Phoenix cluster.  The spectrum is extracted
from a radius of only $1.5$ arcsec centered on the peak of the hard X-ray emission.}
\end{figure*}

We found and intrinsic absorption $NH = (46 \pm 7) \times 10^{22}$ cm$^{-2}$
and unabsorbed intrinsic luminosities of $(3.1\pm 0.1) \times 10^{45}$ and 
$(4.8\pm 0.2) \times 10^{45}$ erg s$^{-1}$ in the $0.5$-$2$ keV and $2$-$10$ keV bands, respectively.
The {\sl Chandra} spectrum of the AGN and the best-fit model are shown in Fig. \ref{AGNspectrum}.
According to this, the central AGN can be classified as a type II QSO.  
These results are consistent with the results of \citet{2013Ueda}, who found 
$NH = (32 \pm 9) \times 10^{22}$ cm$^{-2}$ and $\Gamma = 1.54 \pm 0.27$ from the combined analysis 
of Suzaku XIS and HXD and {\sl Chandra} (with CALDB 4.5.3).  We also note that if we
left $\Gamma$ free in our fit, we found a much flatter spectrum with an intrinsic absorption lower by a factor of 
two, in agreement with the findings of \citet{2013Ueda} for the {\sl Chandra} data alone. This is due to the well-known degeneracy 
between spectral slope and intrinsic absorption.  Given the very low value $\sim 0.6$ for the intrinsic
spectral slope obtained in this way, we prefer to rely on the results with $\Gamma = 1.8$ consistent with the Suzaku+{\sl 
Chandra} analysis.  Since the contribution of the AGN in the soft band is crucial in our spectral analysis
of the core region, we will eventually allow $NH$ to range from $23$ to $53 \times 10^{22}$ cm$^{-2}$, to span the
upper and lower $1\sigma$ limits of the two measurements in the analysis of the XMM data.
Finally, we note that we cannot clearly identify the neutral Fe line at 6.4 keV rest-frame when
adding an unresolved line component.  We also remark that our analysis of {\sl Chandra} data was obtained by removing the 
surrounding ICM emission, and not by modeling the thermal and AGN components together as in \citet{2013Ueda}. 

The total ICM emission contributes about 6400 net counts in the 0.5-7 keV within a radius of about
$650$ kpc, beyond which the surface brightness reaches the background level.  This enabled us
to perform the spectral analysis in independent rings with slightly fewer than 1000 net counts each.
The projected temperature and iron abundance profiles are shown in Fig. \ref{chandraprofiles} in the left and right panels, 
respectively.  The temperature profile clearly shows a prominent
cool core with a decrease of at least a factor of 2 from 200 kpc to the inner 50 kpc.  This is mirrored in the 
iron abundance profile as a clear peak toward the center, where the iron abundance reaches solar metallicity 
\citep[with solar metallicity values measured by][]{2005Asplund} with an uncertainty of 30\%.  These results were used to complement the spectral analysis of  XMM-Newton data.

We also computed the deprojected temperature and density profiles with a backward method \citep[see][]{2010Ettori,2013Ettori},
which makes use of the geometrically deprojected X-ray surface brightness and temperature profiles to reconstruct the hydrostatic mass profile. This is assumed to be
described by the NFW functional form \citep{1996Navarro}.  The best-fit parameters of the mass profile were obtained through the minimization of the $\chi^2$ statistics defined as the sum of 
the squared differences between the observed temperature and the temperature estimates obtained by inverting the equation of the hydrostatic
equilibrium, weighted by the observational errors on the spectroscopic temperature.   The gas density was derived from the deprojected surface
brightness and the total mass model.  With this method we measured a total mass, under the assumption of hydrostatic equilibrium, of $M_{500} = (2.34\pm 0.71) \times 10^{15} \, M_\odot$ at 
$R_{500} = 1627 \pm 235$ kpc.  The ICM mass within $R_{500}$ is 
$M_{ICM}= (2.1 \pm 0.2) \times 10^{14}\, M_\odot$.  From this, we obtained an ICM fraction of
$f_{ICM} =0.09 \pm 0.03$ at $R_{500}$.   As a simple check, we estimated the parameter $Y_X = (27.3 \pm 4.1)\times 10^{14}$ keV $M_\odot$ assuming
as representative of the entire cluster the temperature measured between 100 and 500 kpc, 
and computed the mass estimate from $Y_X$ assuming the relation described in \citet{2010Arnaud} with the slope fixed to the self-similar value.
We found a mass of $(2.15\pm 0.24)\times 10^{15} M_\odot$, consistent with the value previously computed from the hydrostatic equilibrium well within 1 $\sigma$.

If we extrapolate the measured profile up to the virial radius adopting the best-fit NFW profile
 \citep{1996Navarro}, we find $M_{200} = 3.4\pm 1.2 \times 10^{15} \, M_\odot$ at 
$R_{200} = 2500 \pm 400$ kpc.  We note that this value is about a factor of 2 higher than the mass estimate 
$M_{200SZ} = (1.66\pm 0.23_{stat}\pm 0.44_{syst}) \times 10^{15} \, M_\odot$
obtained from SZ-mass scaling relations in \citet{2011Williamson}.  However, considering that SZ-inferred masses 
are found to be statistically lower by a factor of $0.78$ in their sample, the discrepancy between X-ray and SZ mass for the Phoenix 
is reduced to $\sim 1 \sigma$, and therefore is not significant.  Finally, by extrapolating the ICM density distribution, we find 
$M_{ICM}= (3.2 \pm 0.3) \times 10^{14}\, M_\odot$ for a virial ICM fraction of $f_{ICM} = 0.095\pm 0.035$.  

\section{Spectral analysis of XMM-Newton data\label{results}}

\subsection{Analysis strategy}

We extracted the spectra  from a circular region with a radius of
$13.5$ arcsec.  This region was chosen to include the bulk of the emission from the core
region, which is estimated to be confined within a radius of 40 kpc (about $6$ arcsec at $z\sim 0.6$), plus 7.5 arcsec 
corresponding to the half-power diameter of the XMM-Newton PSF.   Since the cold gas is expected to be concentrated toward the
center of the cluster, we assumed that the signal from the cold gas emission outside this radius is negligible.  This assumption is
consistent with the results we obtained from the RGS spectra in two different extraction regions (see Sect. 4.3).

The background was sampled from a nearby region on the same CCD that was free from the cluster
emission and was subtracted from the source spectrum.  We note that the total background 
expected in the source region, computed by geometrically rescaling the sampled background to the 
source area, only amounts to  0.3\% of the total observed emission.   Before performing the final spectral analysis, 
we ran a few  spectral fits artificially enhancing the background by factors of the order 1.2-1.5, 
which range encompasses any possible uncertainty in the background level.  We found no relevant differences in 
the best-fit values as a function of the background rescaling factor.  We conclude that the 
background only weakly affects the spectral analysis. We therefore only consider
the results obtained with the background sampled from the data and geometrically rescaled to the source region.

In XMM-Newton spectra the emission from the central AGN is mixed with the ICM emission,
therefore we modeled its contribution with a power-law spectrum with a fixed slope of 
$\Gamma = 1.8$ and an intrinsic absorption ranging from $23$ to $53 \times 10^{22}$ cm$^{-2}$ at 
the redshift of the cluster, as found in Sect. \ref{spectral_Chandra}.  We also left the normalization 
 free  for each separate spectrum to account for possible differences in the 
 calibration of {\sl Chandra} and XMM-Newton.   We consistently found values in the range $(1.0-1.4)\times 10^{-3}$ for the normalization of the power-law component.
The Galactic absorption, instead, was frozen to the value $NH_{Gal} = 1.52 \times 10^{20}$ cm$^{-2}$ obtained from the radio map of 
\citet{2005LAB} at the position of the cluster, and it is described by the {\tt tbabs} model.  The effects of possible uncertainties on the Galactic absorption are discussed together with other 
systematic effects in Sect. 4.4.
 
We assumed that the gas in the residual cooling flow can be described  by an isobaric cooling model, therefore we used 
the {\tt mkcflow} spectral model \citep{1988Mushotzki}.  This model 
assumes a unique mass-deposition rate throughout the entire temperature range $T_{min}$-$T_{max}$.
Since we are interested in the gas that is completely cooling  and is continuously distributed across the entire temperature range, we set $T_{min} = 0.3$ keV and 
$T_{max} = 3.0 $ keV.   We set the lowest temperature to $0.3$ keV because  this is the
lowest value that can possibly contribute to the emission in the  {\sl Chandra} and XMM-Newton 
energy range, given the relatively high redshift of the Phoenix cluster.  
However, a single {\tt mkcflow} model may not be sufficient to investigate the structure of the cool core.
The actual situation may be more complex, with some of the gas, above a given 
temperature threshold, cooling at a relatively high rate  consistently  with the isobaric
cooling-flow model, while colder gas may have a much lower  mass-deposition rate.
Grating spectra of cool cores are traditionally fitted with an isobaric cooling-flow
model with a cutoff temperature below which no gas is detected \citep{2006Peterson}.
To explore a more complex scenario, we aimed at separately measuring the cooling rate 
(i.e., the mass deposition rate) in several temperature bins.  In this case,
the temperature intervals were fixed to 0.3-0.45, 0.45-0.9, 0.9-1.8, and 1.8-3.0  keV.
This choice is similar to using the single-temperature
{\tt mekal} model for a discrete set of temperatures, but with the advantage of a continuous 
description of the gas instead of a set of discrete temperature values.  

The contribution of the much hotter ICM component along the line of sight
was modeled with a {\tt mekal} model with a free temperature.  Above $3$ keV, 
a single-temperature  {\tt mekal} model can account for several hot components because it is not
possible to resolve the temperature structure above 3 keV with the spectral analysis \citep[see][]{2004Mazzotta}.  This means that  the 
possible contributions to the emission from temperatures between $3.0$ keV (the highest temperature of the 
{\tt mkcflow}) and the best-fit temperature of the {\tt mekal} model are already described by a single-temperature 
{\tt mekal} component.  Therefore, despite the strong temperature gradient toward the center, the 
relevant part of the thermal structure of the ICM is  properly treated by assuming a single temperature for the hot component and 
exploring the low-temperature regime with a multi-temperature {\tt mkcflow} model.

In practice, our fitting method consists of two measurements of the mass 
deposition rates, using the following models:

\begin{itemize}

\item a single cooling-flow model {\tt mkcflow} plus one single-temperature  {\tt mekal} 
component.   The lowest temperature of the {\tt mkcflow} component
is frozen to $0.3$ keV, and the largest to 3.0 keV.  
By setting the lowest temperature to $0.3$ keV, we can interpret 
the normalization of the {\tt mkcflow} model as the global deposition rate allowed
for an isobaric cooling flow across the entire $0.3$-$3.0$ keV temperature range.  
The redshift is tied to the best-fit value found in the hottest component.  This choice does
not introduce any uncertainty given the strong $K_\alpha$ line complex of the H-like and He-like
iron.

\item A set of {\tt mkcflow} models whose lowest and highest temperatures are fixed
to cover the $0.3$-$3.0$ keV range with contiguous and not overlapping intervals.  
Here the normalization of each {\tt mkcflow} component refers to the mass deposition rate
in the corresponding temperature interval.  The upper and lower temperatures are frozen to the following values:
$0.3$-$0.45$, $0.45$-$0.9$, $0.9$-$1.8,$ and $1.8$-$3.0$ keV.  As in the previous case, a 
single-temperature {\tt mekal} model accounts for any gas component hotter than 3 keV.
The redshift is tied to the best-fit found in the hottest component
here as well. 

\end{itemize}

As a fitting method, we used the C-statistics, although we also ran our fits with the 
$\chi^2$-statistics.  In the latter case, spectra were grouped with at least a 
signal-to-noise ratio (S/N) of 4 in each bin.  When running C-statistics, we used unbinned 
spectra (at least one photon per bin).  As a default, we considered the energy range
$0.5$-$10$ keV both for the MOS and the pn data.  We quote the best-fit values with $1 \sigma$ error bars on the measured value of 
$\dot M$, or the 1$\sigma$  upper limit.  The same analysis was also applied to the {\sl Chandra} data
for a direct comparison.

\subsection{Spectral results on the core region from EPIC MOS and pn data}

\begin{figure*}
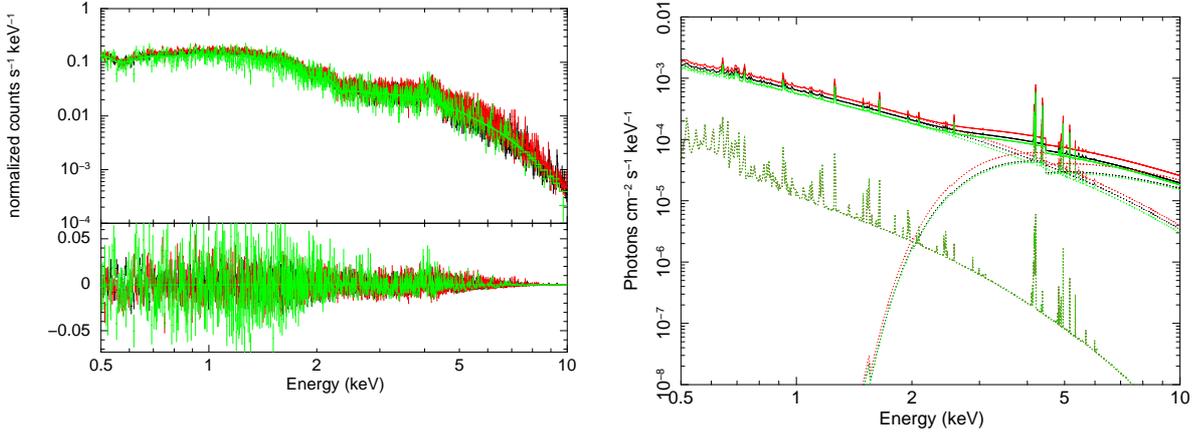

\includegraphics[angle=270,width=8cm]{Phoenix_MOS_mkcflow.ps}
\includegraphics[angle=270,width=8cm]{Phoenix_MOS_mkcflowmodel.ps}
\caption{
\label{MOS_spectra}
{\it Left panel}: MOS spectra from the three Obsids with best-fit model obtained for a single {\tt mkcflow} 
in the 0.3-3.0 keV temperature range.  The lower panel shows the residual with respect to the best-fit model.  {\it Right panel:} components of the best-fit model of the MOS spectra
(three for each component, shown with dotted lines) obtained for a single {\tt mkcflow} in the 0.3-3.0 keV temperature range.  
The {\tt mkcflow} component is the lower thermal component, while the AGN contribution is virtually negligible below 2 keV.}
\end{figure*}

\begin{figure*}
\includegraphics[angle=270,width=8cm]{Phoenix_pn_mkcflow.ps}
\includegraphics[angle=270,width=8cm]{Phoenix_pn_mkcflowmodel.ps}
\caption{
\label{pn_spectra}
{\it Left panel}: pn spectra from the three Obsids with best-fit model obtained for a single {\tt mkcflow} 
in the 0.3-3.0 keV temperature range.  The lower panel shows the residual with respect to the best-fit model.   {\it Right panel:} components of the best-fit model of the pn spectra
(three for each component, shown with dotted lines) obtained for a single {\tt mkcflow} in the 0.3-3.0 keV temperature range.  
The {\tt mkcflow} component is the lower thermal component, while the AGN contribution is virtually negligible below 2 keV.}
\end{figure*}

\begin{figure*}
\includegraphics[width=8cm]{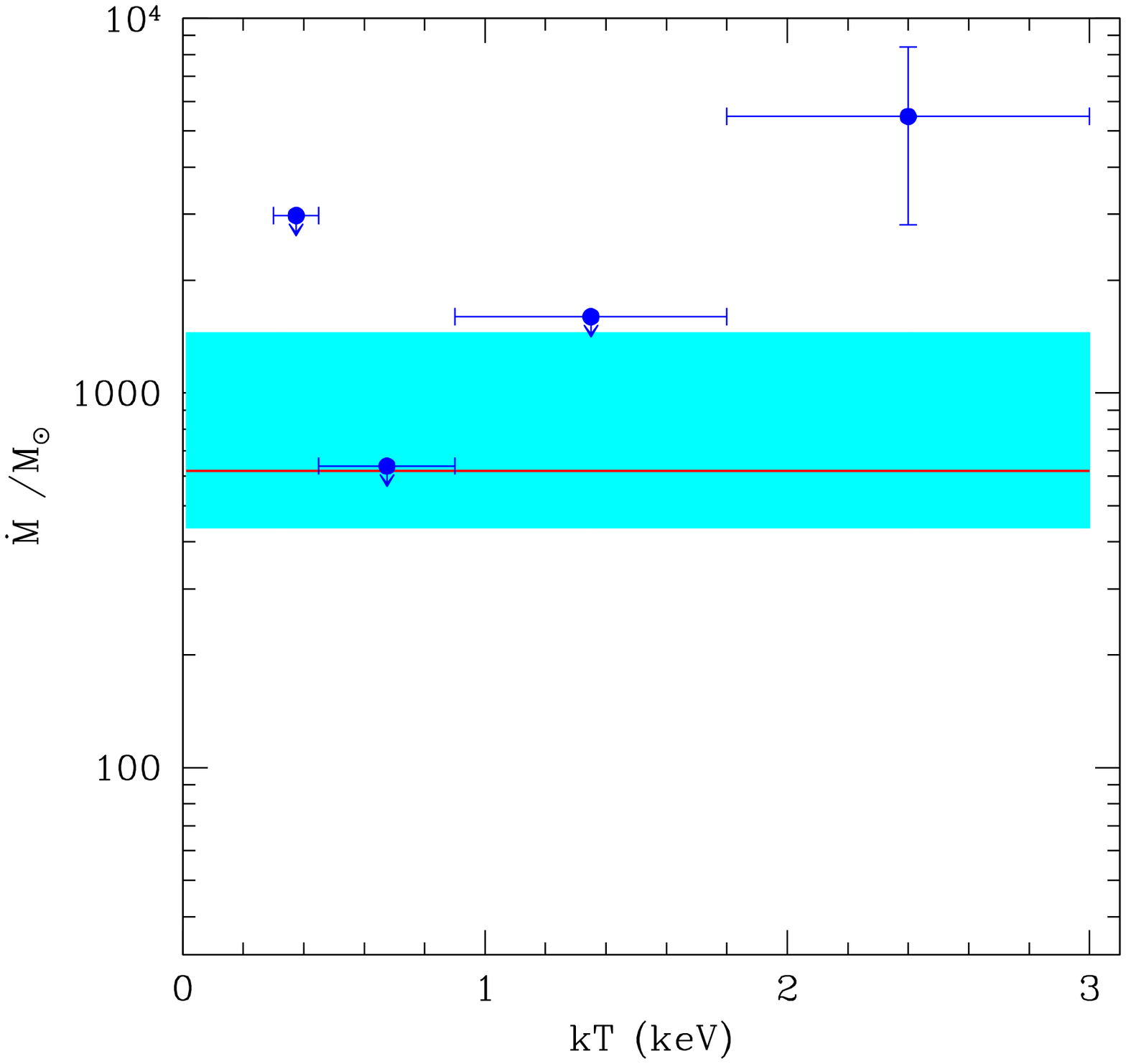}
\includegraphics[width=8cm]{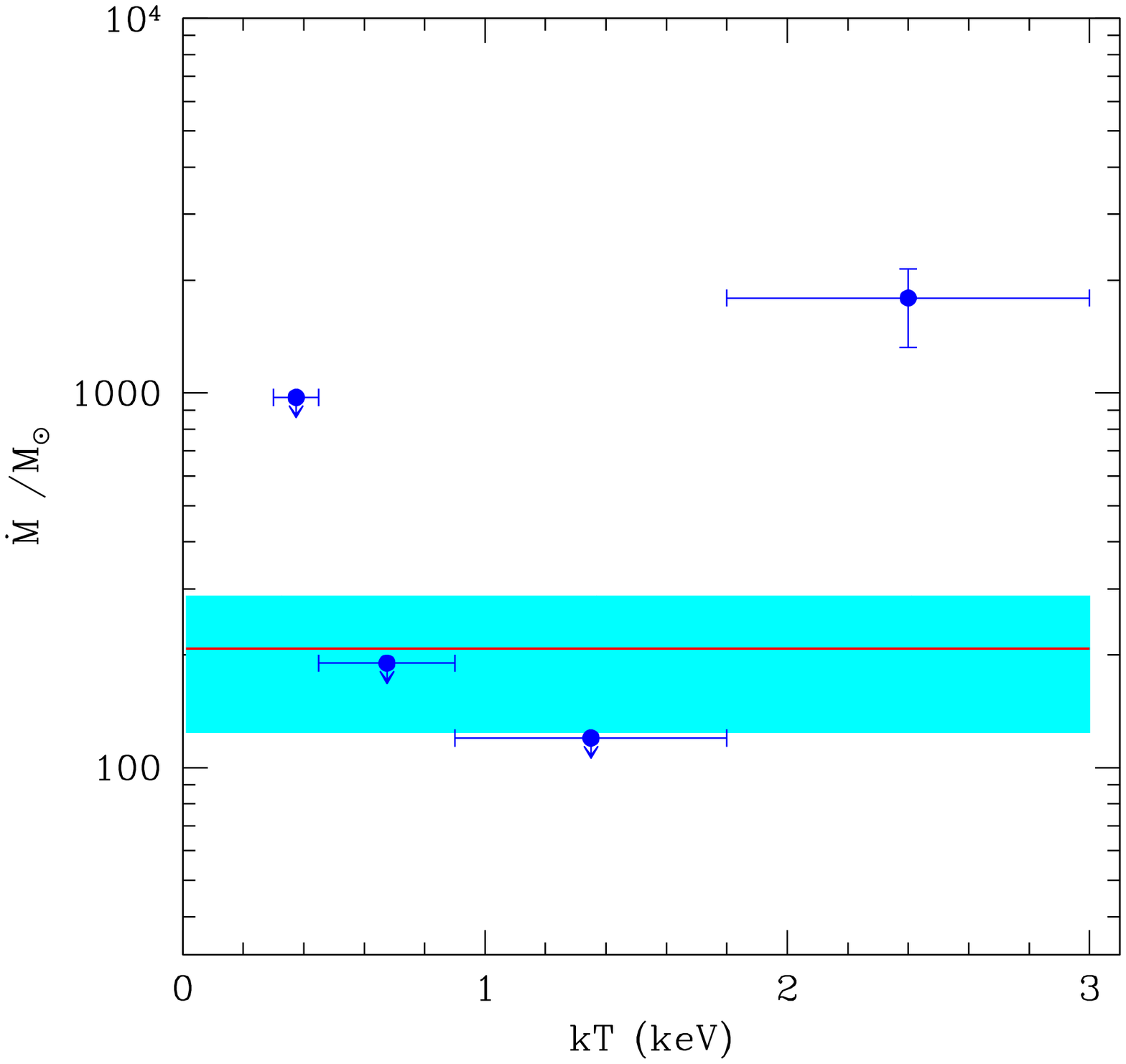}
\caption{
\label{mdot_mkcflow}
{\it Left panel}: mass deposition rate $\dot M$ as a function of the temperature
from the analysis of the MOS data of the Phoenix with a multi-component {\tt mkcflow} model. 
Error bars are at 1 $\sigma$ confidence level and arrows represent upper limits at 1 $\sigma$ (the systematic uncertainty is not included here).   
The red horizontal line shows the best-fit value from the single  {\tt mkcflow} model in the 0.3-3.0 keV temperature 
range, while the shaded area shows the  1 $\sigma$ uncertainty.
{\sl Right panel:}  Same as in the left panel, but for the pn data. }
\end{figure*}

Before fitting the XMM data, we performed a series of basic tests on our spectra.  First, we separately fit pn and MOS spectra
from each exposure using a single {\tt mekal} model plus the fixed AGN component and Galactic absorption.  
The best-fit temperature values were all consistent within $1\, \sigma$, showing that 
all the spectra are broadly consistent with each other.  When we added a {\tt mkcflow} component, we noted that the best-fit values
of the mass deposition rates from pn data were  lower than those from MOS data in all the different exposures.  This shows that there may be significant differences between  pn and MOS
calibration that affect the measurements of cold gas components.  Therefore, we proceeded with our standard analysis strategy,
which combines different exposures of the same detector, but fitted  pn and MOS spectra  separately.

Following our analysis strategy, first we separately fitted the MOS and pn data with a 
single {\tt mkcflow} model, coupled with a {\tt mekal} model to account for the hot ICM along 
the line of sight.  We used both our new XMM-Newton data and the archival data.  We considered
the energy range 0.5-10 keV  both for MOS and pn data.  The \xspec \, version used in this work is v12.8.1.

First we used a single {\tt mkcflow} model with $T_{min}=0.3$ keV and $T_{max} = 3.0 $ keV to fit
the combined MOS data with Cash statistics. The redshift of the cold gas component is linked to the best-fit redshift 
found for the hot component (described by a {\tt mekal} model) 
$z_X = 0.593 \pm 0.002$, consistent with the optical value \citep{2012McDonald}.
We found that the metal abundance of the {\tt mkcflow} component is not constrained by the
present data.  Therefore we chose to explore the best-fit mass deposition rate by varying the cold
gas metallicity $Z_{cold}$ in a wide range of values.   We set the lowest value for
$Z_{cold}$ equal to the metallicity measured for the hotter {\tt mekal} component 
$Z_{hot} = 0.47_{-0.03}^{+0.04} Z_\odot$ \citep[with solar metallicity values measured by][]{2005Asplund}, and we set the upper limit to the 
$2\sigma$ upper value  $\sim 1.4 Z_\odot$ found in the inner 80 kpc in the {\sl Chandra} analysis.
We quote the best-fit values obtained for $Z_{cold} = 1.0 \, Z_\odot$, and add the uncertainty 
associated to the range of $Z_{cold}$ as a systematic error.  We found a global mass deposition rate
$\dot M =  620\, \,  (-190\, \, +200)_{stat} \, \, \, (-50 \, +150)_{syst} \, M_{\odot}$ yr$^{-1}$.  
The intrinsic absorption of the AGN is found to be $NH = (42.0 \pm 1.5) \times 10^{22}$ cm$^{-2}$, 
in excellent agreement with the value found with the {\sl Chandra} analysis.
The best-fit temperature of the hot component is $ kT_{hot}= (6.5 \pm 0.3) $ keV, which also agrees well with the
emission-weighted temperature in the inner 80 kpc obtained from the analysis of the {\sl Chandra} data.
We repeated the fit with $\chi^2$-statistics  with spectra binned to at least an S/N$=4$ per bin, 
and we found almost identical results, with a reduced $\chi^2 = 1.04$ for 663 dof.  
The binned spectra of the three Obsids with the best-fit models are shown in Fig. \ref{MOS_spectra}, left
panel.  We show the different components of the best-fit
model in the right panel of this figure.  We note that the contribution of the cold gas in the temperature range 0.3-3 keV is 
about 5\% in the 0.5-1 keV energy band, while the strongly absorbed AGN has virtually no emission
below 2 keV, also when assuming the maximum uncertainty in the intrinsic absorption value.
We conclude that the MOS data provide a positive detection of an average mass-deposition rate
in the temperature range 0.3-3.0 keV at the $3\sigma$ c.l., with
$\dot M > 190 M_\odot$ yr$^{-1}$ at the $2\sigma$ c.l. after considering the systematic effect associated
with the unknown metallicity of the cold gas.  For completeness, we repeated our analysis without merging MOS1 and
MOS2 spectra, {which implies a combined fit of six independent spectra}.  We found similar results and error bars for all the parameters, with the best-fit values
for the mass deposition rate lower by 6.5\%.

Then we ran the multiple {\tt mkcflow} model on the combined MOS spectra.  The constraints on the amount of cold gas vary significantly as a function of the 
temperature range.  In this case, the metal abundance of the cold
gas is linked to the value found in the temperature range 1.8-3.0 keV, which is $Z_{1.8-3 keV} = 
0.36_{-0.15}^{+0.19}\,  Z_\odot$.  The best-fit temperature of the {\tt mekal} component is now
$kT_{hot} = 7.4_{-0.4}^{+0.5}$ keV with a metal abundance of $Z_{hot} = 0.54_{-0.04}^{+0.05} \, Z_\odot$. 
The results for $\dot M$ are shown in Fig. \ref{mdot_mkcflow}, left panel.  We note that the lowest upper limits (at $1 \sigma$) 
are measured in the temperature bins $0.45-0.9$ and $0.9-1.8$ keV.  These upper limits provide the 
strongest constraint for the fit with the single {\tt mkcflow} model (see shaded area in Fig.  \ref{mdot_mkcflow}).  We also note that a clear detection of a mass deposition rate
is obtained for the 1.8-3.0 keV temperature range.   This means
that our analysis of the MOS data with a temperature-dependent {\tt mkcflow} model 
suggests that the cold gas around and below 1 keV does not cool
as rapidly as the gas between 2 and 3 keV.  This is consistent with the physical conditions often encountered in cool cores, where
the gas cools down to a temperature floor below which it is hard to probe the presence of cooling gas.  However, in the case of the Phoenix, the gas is observed to cool down to
a temperature much lower than the ambient temperature $T_{vir}$, as opposed to the classic cool cores where the lowest temperature is $\sim 1/3 T_{vir}$.  
We only obtained upper limits on $\dot M$ for gas with temperatures below 2 keV.  From comparing the results
from the multi-temperature {\tt mkcflow} model with that from the single-temperature {\tt mkcflow}, we conclude that the mass deposition 
rate found in the analysis with a single {\tt mkcflow} model should not be taken as representative of the entire cooling flow, since it
is an emission-weighted value averaged over a temperature range that is too wide.  Therefore, 
a temperature-resolved analysis of the cool core structure is mandatory to constrain the
presence of a cooling flow.   Nevertheless, the constraints on the mass deposition rate from MOS data 
still allow values of several hundreds of $M_\odot$ yr$^{-1}$, 
which might agree with the star formation rate observed in the BCG.  Finally, the same fit on the 
separate MOS1 and MOS2 spectra provides very similar upper limits.  In particular, the most constraining bin at $0.45<kT<0.90$ keV is left unchanged.

The fit of the pn data with the single {\tt mkcflow} model encountered some difficulties.  The best fit was obtained 
when the temperature of the hot component was $kT_{hot}=5.9 \pm 0.1$ keV, which 
disagrees with the MOS data and with the {\sl Chandra} results.  
In addition, the reduced $\chi^2 = 1.38$ for 637 degrees of freedom, is significantly larger than the
value obtained for the MOS data.  The lower temperature of the hot component is enough to account
for most of the soft emission, which in the MOS fit was associated with the cold gas.  As a result, the
soft emission that can be associated with the cold gas is significantly lower, and the $3\sigma$ 
upper limit to the mass deposition rate is as low as $\sim 240 M_\odot$ yr$^{-1}$.  

The significant difference between pn and MOS results clearly requires a detailed treatment of 
any cross-calibration uncertainty, which is beyond the goal of this paper.  A first attempt to estimate the
systematic uncertainty due to cross-calibration problems is discussed in Sect. 4.4.
Nevertheless, a reliable assumption which may bring the two analyses into better agreement is to set the temperature of the hot gas to the value
found with  {\sl Chandra} and XMM-Newton  MOS, 
 $kT_{hot} \sim  6.8$ keV.    With this assumption, we find $\dot M =  210\, \,  (-80 \, \, +85)_{stat} \, \,   
 (-35 \,\, +60)_{syst}\, M_\odot$ yr$^{-1}$.  
Here, the best-fit value was also obtained after freezing the abundance of the cold gas to 
$Z_{cold} = 1.0 \, Z_\odot$, and  the systematic error was obtained
by varying  the abundance of the cold 
gas in the range  $0.45$-$1.4\,  Z_\odot$.  The abundance of the hot {\tt mekal} component is
$Z_{hot}=0.48\pm 0.04 \, Z_\odot$ , which agrees very well with the MOS data.
Finally, the intrinsic absorption of the AGN is also well constrained to be
$NH = (43.6 \pm 1.2) \times 10^{22}$ cm$^{-2}$.  
The binned spectra of the three Obsids with the best-fit models are shown in Fig. \ref{pn_spectra}, left
panel.  In Fig. \ref{pn_spectra}, right panel, we show the different components of the best-fit
model. It is possible to appreciate the lower {\tt mkcflow} component with respect to the right panel of
Fig. \ref{MOS_spectra}.   This means that the analysis of the  pn data plus the condition $kT_{hot} = 6.8$ keV 
provides a positive detection of a significant mass-deposition rate, although about a factor of 3
lower than those found in the MOS analysis.  

We briefly comment on the dependence of $\dot M$ on the temperature of the surrounding hot gas.  
Clearly, the value of $\dot M$ critically depends on the amplitude of the continuum and
therefore on the hot gas temperature.  As already mentioned, our strategy consisted of anchoring the hot gas temperature to the best-fit value driven by the 
continuum in the high-energy range, while the colder gas below 3 keV was properly treated by the multi-temperature {\tt mkcflow} model.
With current data, any other treatment of the hot gas component would be highly speculative, therefore we did not explore the possible effect of 
the temperature distribution above 3 keV in the cluster core
in more detail.  We argue that the best way to deal with this aspect is to use highly spatially resolved data
that allow avoiding the surrounding hot gas component as best
possible and to focus on the innermost core region.

The analysis with a multi-temperature {\tt mkcflow} model, shown in the right panel of Fig. \ref{mdot_mkcflow}, confirms this inconsistency.  
The behavior of $\dot M$ as a function of the temperature is similar, 
with the most constraining upper limits coming from the temperature bins  $0.45-0.9$ and $0.9-1.8$ keV and a clear detection of a high mass-deposition rate
in the 1.8-3.0 keV temperature bin.   The ratio between $\dot M$ in the bin $1.8-3.0$ keV and the upper limits below 1.8 keV also has the same 
value as in the MOS analysis.  We therefore qualitatively reach the same conclusion as was obtained from the MOS data analysis with the multi-temperature
{\tt mekal} model, but the values  $\dot M$ are scaled down by a factor $\sim 3$.  Incidentally, if the energy range used for spectral fitting
is reduced to the 0.7-10.0 keV range, best-fit values and upper limits from the pn spectra are found to agree with MOS results.  Clearly, this simply depends on the 
fact that the sensitivity to the cold gas is significantly reduced by excluding the energy bins below 0.7 keV, allowing much higher upper limits to the contribution from the cold gas.

To summarize, the analysis of XMM-Newton CCD spectra confirms the detection of gas cooling at a rate 
$>1000 M_\odot $ yr$^{-1}$ in the temperature range 1.8-3.0 keV, while it only provides upper limits for 
gas at temperatures $kT < 1.8$ keV.
The upper limits to the global mass deposition rate below 1.8 keV appear to be consistent with an SFR of 
$\sim 800 M_\odot$ yr$^{-1}$ within $1\sigma$, as measured by \citet{2013aMcDonald} in the BCG, from the MOS data analysis.
On the other hand, the upper limits on $\dot M$ measured from pn data at temperatures below 1.8 keV are
significantly lower (more than $3\sigma$) than the SFR.  Therefore, no final conclusion can be drawn on the 
correspondence between the cooling flow  and the observed star formation rate in the core of the Phoenix cluster.

\subsection{Spectral results from RGS data}

The RGS spectrum was extracted from a strip of roughly 50 arcsec long across the center of the object, 
a width corresponding to 90\% PSF.  Because the RGS is a slitless instrument 
and the zero-point of the wavelength calibration
is dependent on the position of the detector in the field of view, we used as the source center the coordinates
$RA=23$:$44$:$43.9$, $DEC=-42$:$43$:$12.64$, which correspond to the position of the central AGN (see Sect. 3).
In practice, with our choice for the extraction region, we collect photons from a box 
region with dimensions $50" \times 12'$ centered on the source \citep[see, e.g., Figs. 2 and 3 in ][]{2009Werner}.
This is the standard choice to maximize the signal from the cluster.  
This choice is motivated also by the fact that the Phoenix cluster has a brightness distribution that is
similar to a point source from the XMM point of view, so that the line widening that is due to the spatial extension
of the source does not severely affect the data.  
This means that the effective area that is sampled by our RGS spectrum is a circle with a radius of 25", 
to be compared with the circle of 13" used to extract the EPIC spectra.   We also extracted an RGS spectrum from 
a narrower region with dimensions $15" \times 12'$, corresponding to 70\% of the PSF as opposed 
to the 98\% of the PSF achieved with a width of $50"$.  

\begin{figure*}
\begin{center}
\includegraphics[width=12cm]{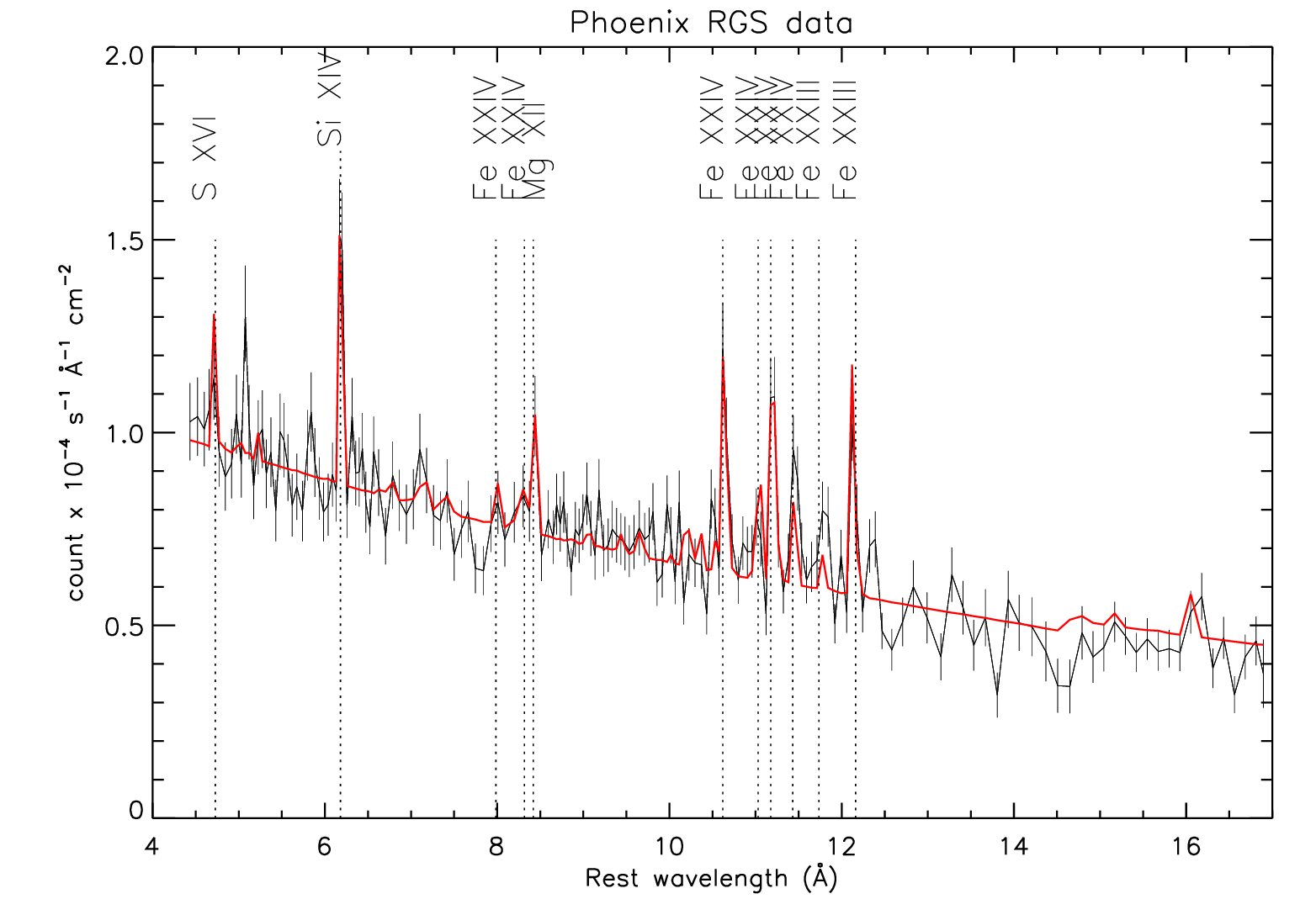}
\includegraphics[width=12cm]{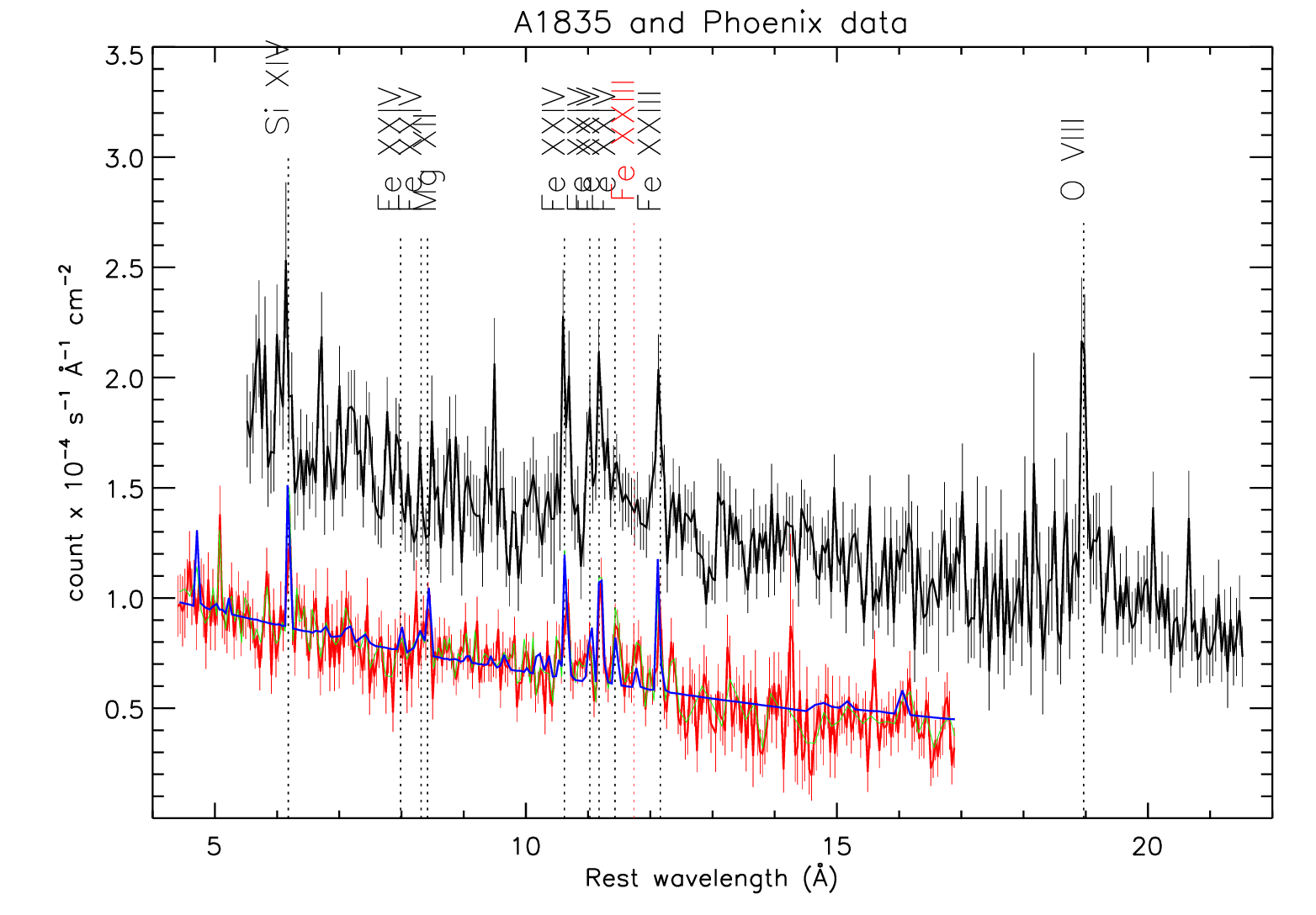}
\end{center}
\caption{
\label{fig.spectra}
{\it Top panel}: combined RGS 1 and 2 spectrum for the Phoenix cluster and best-fit single-temperature model.
The data have been rebinned to have a signal-to-noise ratio of 10 and divided by the effective area of the instrument only for displaying purpose.  Relevant lines are also labeled.
{\it Bottom panel}: The Phoenix spectrum (in red) is compared with the RGS spectrum of Abell 1835 (in black). We plot the fluxed spectra obtained with RGSFLUXER.  Similar results are obtained by plotting 
the unfolded spectrum in \xspec \, (the green line in the Phoenix spectrum) wit the only difference set by the binning condition $S/N = 10$. The best-fit thermal model of the Phoenix is also plotted as a blue line. 
We also plot relevant lines clearly seen in the spectra: the same sets of lines are present in both spectra in the overlapping wavelength range.}

\end{figure*}

We fitted the first-order spectra between 7 and 27 $\AA$, since this is the range where the source is higher than the background.
We used \xspec \, version 12.8.1 and C-statistics on the unbinned spectrum.  We first applied a single-temperature {\tt mekal}
model modified by the Galactic absorption, while the redshift was fixed at the optical value.
The AGN was modeled as in the fits of the EPIC data, but because
of the strongly absorbed spectrum, the effect 
of the AGN emission is completely negligible in the RGS spectra given the limited wavelength range.  
From the RGS spectrum extracted from the larger regions (with a width of $50"$),
we found that a single-temperature model provides a reasonable fit to the data, with a 
$C_{stat}$/dof = 1978/1952 with a best-fit temperature of $kT = 6.2\pm 0.4$ keV with a metal abundance of $Z = 0.55 \pm 0.09\, Z_\odot$.
We show in the left panel of Fig. \ref{fig.spectra} the first-order combined spectrum with the best-fitting thermal model.  In the same figure we show the
positions of relevant emission lines that are clearly identifiable in the spectrum.  The set of lines is the same seen in the RGS spectrum of the 
cluster Abell 1835, as shown in the comparison plot in the bottom panel of Fig. \ref{fig.spectra}.  For Abell 1835 we used the same data reduction as performed on the
Phoenix data on the three observations not affected by flares (Obsid 0098010101, 0551830101, and 0551830201).  The spectrum obtained with our reduction agrees
very well with the spectrum shown by \citet{2010Sanders}.  Based
on the RGS spectrum of the Phoenix cluster we therefore reach the same conclusions as were obtained for A1835:
no line emission from ionization states below Fe XXIII is seen above $12 \AA$, and no evidence for gas cooling below $\sim 3$ keV is found.    

We then proceeded to obtain a measurement of the mass deposition rate from the RGS data by using the {\tt mekal+mkcflow} model with abundances constrained to be the same in
both components.  As for the EPIC data analysis, the temperature range of the {\tt mkcflow} component was fixed to $0.3-3.0$ keV.  
We found no statistically significant improvement with respect to the single-temperature model ($C_{stat}$/dof = 1977/1951).  
The best-fit ambient hot temperature was $kT_{hot} = (6.3\pm 0.5)$ keV with a metal abundance of $Z = 0.43 \pm 0.09\, Z_\odot$.  The best-fit value for the mass deposition rate was
$\dot M = 122_{-122}^{+343}$ $M_{\odot}$ yr$^{-1}$.  The 90\% upper limit on the mass deposition rate was $682 \, M_{\odot}$ yr$^{-1}$. 

More complicated models, such as the one including a set of {\tt mkcflow} models, are not constrained by the RGS data.  However, we were able to 
obtain constraints on the abundance of metals other than iron using a {\tt vmekal} model.  We found 
$Z_{Mg} = (0.53\pm 0.24) \, Z_\odot$, $Z_{Si} = (0.66\pm 0.15) \, Z_\odot$, and $Z_{S} = (0.33\pm 0.25) \, Z_\odot$.   The use of {\tt vmekal} 
provides a slightly lower upper limit on the mass deposition rate: $\dot M = 103_{-103}^{+345}$ $M_{\odot}$ yr$^{-1}$, with a 90\% upper limit 
of $590 \, M_{\odot}$ yr$^{-1}$.  Best-fit temperatures are left unchanged.

We applied the same analysis to the RGS spectrum extracted from a smaller region with a width of $15"$.  
We obtained very similar results, with small differences, the largest being a lower value of about 1 $\sigma$ of the 
ambient hot temperature, which was meaured to be $kT_{hot} = 5.7^{+0.6}_{-0.3}$ keV.  As for the mass deposition rate, 
we found a 1 $\sigma$ upper limit of $\dot M = 470\,  M_\odot$ yr$^{-1}$, with a  90\% upper limit at 
$744  \, M_\odot$ yr$^{-1}$.   Therefore, the RGS spectral analysis provides similar results for the two different extraction regions.

To summarize, the analysis of the RGS data only provided upper limits to the global mass deposition rate, in agreement with the values 
found with the analysis of the EPIC pn data, and also consistent with those found with EPIC MOS within less than $2 \sigma$ .  The better agreement with EPIC pn analysis is expected on the basis of  
the cross-calibration status of XMM instruments:  the agreement between EPIC pn and RGS is currently within a few percent as a result of the combined effect of the RGS contamination model plus the  improved pn
redistribution model \citep{2010Stuhlinger}.

At the same time, we remark that the comparison between the
values of $\dot M$ obtained from the two RGS extraction regions and
the EPIC pn extraction region allows us to conclude that
our choice of the extraction radius for EPIC data was adequate for sampling the possible cold gas emission.  The simple fact that $\dot M$ from RGS  sampled
in the region twice larger than that used for EPIC data is not significantly larger than the value found in the pn data
confirms that no significant emission from cold gas can be found outside the 13" radius.

\subsection{Discussion of the systematics in XMM-Newton data analysis}

The discrepancy among the results obtained with MOS and pn can be ascribed to the uncertain cross-calibration
of the effective area among the XMM-Newton CCD, while there are no other sources of statistical noise
that can account for a significant fraction of this discrepancy.  This difference is persistent despite the recent model for the 
MOS contamination introduced in the SAS release v13.5.0 \citep[see][]{2013Sembay}.  
Since the contribution of the cold gas below 3 keV to the total emission in the extracted region is about 3\% in the 0.5-2.0 keV energy range and
about 5\% in the 0.5-1.0 keV range, an uncertainty in the effective area of the same order may 
severely affect the measurement of the cold gas.  

A first way to estimate the calibration uncertainty is to compare our results with the results obtained by applying the correction CORRAREA to the response area of the EPIC detectors 
based on the results of \citet{2014Read}.  The CORRAREA calibration  is based on an extensive
cross-calibration study of 46 non-piled-up sources extracted from the 2XMM EPIC Serendipitous Source Catalogue \citep{2009Watson}.  
This phenomenological correction is meant to bring into agreement the broadband fluxes measured by EPIC-MOS and EPIC-pn.  
We obtained an estimate of the uncertainty on $\dot M$ values associated to uncertainties in the EPIC response areas by comparing the best-fit values
obtained with and without the CORRAREA correction.  The best-fit values of $\dot M$ are 1\% and 5\% higher when applying the CORRAREA 
correction for MOS and pn data, respectively.  This is far from the factor of 3 needed to bring $\dot M$ from MOS and pn into
agreement.  Therefore
we conclude that the measurement of cold gas in the spectra of the Phoenix cluster is significantly affected by the calibration of the EPIC detectors at a level 
beyond that probed by \citet{2014Read}.

We also included the energy range 0.3-0.5 keV to investigate possible effects associated with this low-energy band.  
Although the calibration is even more uncertain in this range, the inclusion of the lowest energies may help
to increase the relative contribution of the cold gas.  However, when we repeated our fits on the 0.3-10 keV 
energy range, we did not find any significant difference with respect to the results described in Sect. 4.2.  

We also explored the possibility that data taken in different epochs may have a different calibration.  However, when we
excluded the shortest Obsid, which was acquired two years earlier than most of our data on the Phoenix, the best-fit
values were only affected by a negligible amount.

A main source of uncertainty in the soft band is the Galactic column density.  The possible presence of 
unnoticed fluctuations in 
the Galactic neutral hydrogen column densities on scales smaller than the resolution of 
\citet{2005LAB} may reach 10-40\%  on  scales of $\sim $ 1 arcmin \citep{2003Barnes}.  
Therefore,  we considered a systematic uncertainty in our best-fit values of $\dot M$ assuming
a maximum variation of $NH_{Gal}$ by 40\%.  The typical effect is that the best fit is obtained for the highest
allowed value of $NH_{Gal}$, and this implies that much more cold gas can be allocated.  Specifically, if we 
allow the Galactic absorption to vary up to $NH_{Gal} = 2.13 \times 10^{20}$ cm$^{-2}$, 
we find  $\dot M =  900 (-390\, \, +110)_{stat} ( -130\, \, +250)_{syst} M_{\odot}$ yr$^{-1}$
from the analysis of the MOS data with a single {\tt mkcflow} model in the temperature range 0.3-3.0 keV.
This value is 45\% higher than that obtained with the $NH_{Gal}$ value of  \citet{2005LAB} at the position of the cluster.  
A similar effect is found for the other fits, including those with the multi-component {\tt mkcflow} model.  
If, on the other hand, we leave $NH_{Gal}$ free to vary, we find best-fit values of $NH_{bestfit} = 
(2.2 \pm 0.8) \times 10^{20}$ cm$^{-2}$, which is very close to the upper bound we assumed for $NH_{Gal}$.  
The analysis of the EPIC pn data, with the same values of $NH_{Gal}$, provides $\dot M$ values 2.3 and 3 times higher
for $NH_{Gal} = 2.13$ and $2.2 \times 10^{20}$ cm$^{-2}$, respectively, which considerably reduces the discrepancy between the 
MOS and pn analysis.  However, when the Galactic absorption is left free, the best-fit value is $NH_{Gal} = (0.8 \pm 0.2) \times 10^{20}$ cm$^{-2}$, 
which in turn  provides much lower values of $\dot M$ than the standard analysis.  Therefore, we conclude that there are no hints for a plausible variation of  $NH_{Gal}$ 
that can significantly change our results, including the
discrepancy between the best-fit values of $\dot M$ found between MOS and pn data.

\subsection{Comparison with spectral results on the core region from {\sl Chandra} data}

\begin{figure}
\begin{center}
\includegraphics[width=8cm]{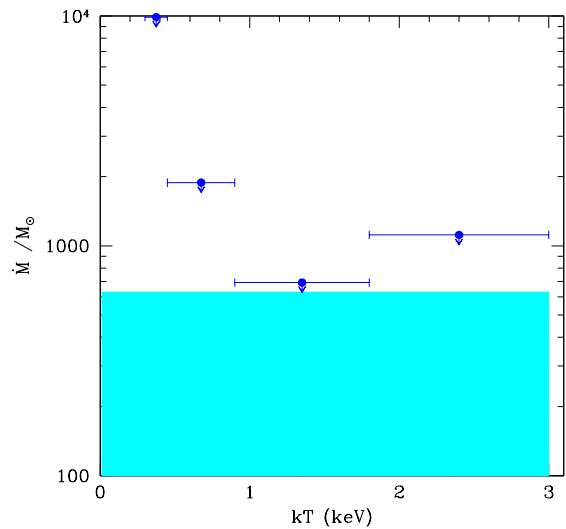}
\caption{\label{mdot_Chandra}
Mass deposition rate $\dot M$ as a function of the temperature
from the analysis of the ACIS-I {\sl Chandra}  data of Phoenix within 30 kpc
from the center.  The AGN emission has been removed (inner 1.5 arcsec).  
The results have been obtained with a multi-temperature {\tt mkcflow} plus a {\tt mekal} model for the 
hot component.  Arrows refer to 1 $\sigma$ upper limits.  The
shaded area shows the 1 $\sigma$ upper limit based on the analysis
of a single-temperature {\tt mkcflow} model in the 0.3-3.0 keV temperature range plus a {\tt mekal} model for the hot component.}
\end{center}
\end{figure}

We repeated the same fits on the {\sl Chandra} data available in the archive as of December 2014.  
The photometry in the inner 30 kpc, corresponding to 4.5 arcsec, excluding the central AGN, amounts to 1510 net counts in the 
$0.5-7.0$ keV energy band.  This is usually not sufficient to provide robust constraints on the amount 
of cold gas in nearby clusters, and it would be 
even less effective at the redshift of Phoenix.  Nevertheless, we performed our spectral analysis
as for the XMM-Newton data.  The single-temperature {\tt mkcflow} model in the temperature range 0.3-3.0 keV provides a 
1 $\sigma$ upper limit of $630 \, (-85 \,\, +60)_{syst}\,\,    M_{\odot} $ yr$^{-1}$ or a $2 \sigma$ upper limit of 
$1130  \, (-150 \,\, +1100)_{syst}\,\, M_{\odot} $ yr$^{-1}$, where the systematic uncertainty corresponds to
$0.45 < Z_{cold} < 1.4\,  Z_\odot$.  The hot gas temperature is $kT_{hot} = 7.2_{-0.6}^{+1.3}$ keV, within 
$1 \sigma$ from the value found in XMM-Newton analysis and in the overall temperature profile of the 
{\sl Chandra} data.  We recall that we did not fit the AGN since its emission was removed from 
the spectrum thanks to the {\sl Chandra} angular resolution.

Despite the low S/N of the {\sl Chandra} data  (almost two orders of magnitude fewer photons than for 
the combined XMM-Newton spectra), and the high redshift of the Phoenix cluster, we also performed the fit
with a multi-component {\tt mkcflow} model.  Temperature and metallicity of the hot
gas component were the same as in the previous fit, while $Z_{cold}=1.0 \, Z_\odot$.  
As expected, we obtained little additional information.
The results are summarized in Fig. \ref{mdot_Chandra}, where we can conclude that the 
strongest constraints on $\dot M$ mostly comes from the gas between 0.9 and 1.8 keV.  
This means that  {\sl Chandra} upper limits are consistent with a mass deposition rate of about $\sim 1000 M_\odot$ yr$^{-1}$.  
We remark, however, that deeper {\sl Chandra} data are expected to provide much stronger constraints.

\section{Exploring the connection between SFR and global $\dot M$\label{rev_SFR}}

\begin{figure}
\begin{center}
\includegraphics[width=8cm]{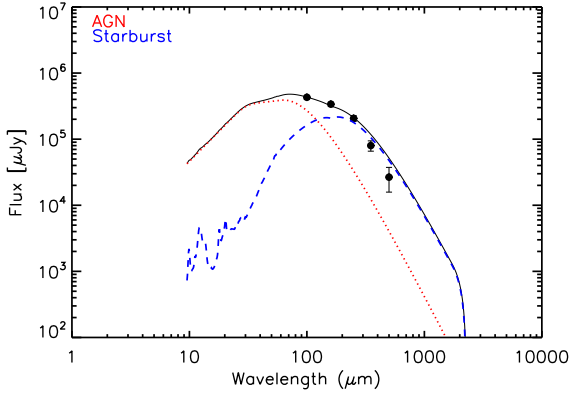}
\caption{\label{FIR_spec}Spectral energy distribution of the BCG of Phoenix in the FIR from the PACS (100 and 160 $\mu$m) and 
SPIRE (250, 350, 500 $\mu$m) instruments onboard the   {\sl Herschel} Space Observatory. The red dotted line and the blue dashed line
show the best fit to the SED for the AGN and starburst template, respectively.}
\end{center}
\end{figure}

We inspected available FIR data on the Phoenix cluster to better constrain the SFR.
The BCG has been observed in the FIR with the PACS (100 and 160 $\mu$m) and 
SPIRE (250, 350, 500 $\mu$m) instruments onboard the   {\sl Herschel} Space Observatory \citep{2010Pilbratt}. 
  {\sl Herschel} data bracket the critical peak of FIR emission of high-redshift galaxies, providing a 
direct, unbiased measurement of the dust-obscured star-formation.  We performed aperture photometry 
assuming an aperture radius of 6 and 9 arcsec for 100 and 160 $\mu$m, respectively, while for SPIRE we used PSF fitting.
The five data points from  {\sl Herschel} instruments provide the SED 
shown in Fig. \ref{FIR_spec}.  

We investigated the contribution of the AGN component to the FIR emission using the program DECOMPIR \citep{2011Mullaney}, an
SED model-fitting software that aims to separate the AGN from the host star-forming (SF) galaxy.  The AGN component is
an empirical model based on observations of local AGNs, whereas the five starburst models were developed to
represent a typical range of SED types, with an extrapolation beyond
100 $\mu$m using a gray body with emissivity $\beta$ fixed to 1.5.  
The best-fitting model obtained with DECOMPIR confirms 
at least a 50\% contribution of the AGN to the total SED flux, as shown in Fig. \ref{FIR_spec}.
The AGN clearly dominates the FIR emission.  The IR luminosity for the starburst component
\citep[best described by the model SB5 in][]{2011Mullaney}
is measured to be $L_{IR} =  3.1 \times 10^{12}\, L_\odot$.  Assuming a Salpeter IMF, 
this luminosity corresponds to a star formation rate of $530 \, M_\odot$ yr$^{-1}$  with a typical error of 15\%, which is
lower than the value found by \citet{2013aMcDonald} at 3 $\sigma$ c.l.   
The total luminosity given by the best fit is $L_{IR} =  2.09 \times 10^{13}\, L_\odot$, with the AGN component contributing 
by 86\%.   Taken at face value, this revised estimate of the SFR  is in very good agreement with the 
mass deposition rate found with the analysis of the EPIC MOS data, while is still inconsistent at 
about $3 \sigma$ c.l. with the mass deposition rate found with EPIC pn analysis.

\section{Conclusions\label{conclusions}}

We analyzed the X-ray data taken with a 220 ks exposure of
XMM-Newton on the Phoenix cluster.   We focused on cold gas in the core by selecting 
a circle of $13.5$ arcsec centered on the BCG in the XMM image and a strip with a width corresponding to 90\% of the
PSF in the RGS data.  Our immediate goal was  to constrain the actual mass-deposition rate associated with the residual cooling flow, with the aim of 
understanding whether this may be directly linked to the current massive starburst
observed in the BCG.  We combined XMM-Newton data with shallow {\sl Chandra} data, particularly to model the
hard emission of the central absorbed AGN, which cannot be removed from the XMM-Newton data alone.
Our results are summarized as follows:

\begin{itemize}

\item We measured an average mass-deposition rate of 
$\dot M =  620\, \,  (-190\, \, +200)_{stat} \, \, \, (-50 \, +150)_{syst} M_{\odot}$ yr$^{-1}$
and $\dot M =  210 \, \,  (-80 \, \, +85)_{stat} \, \,   
 (-35 \,\, +60)_{syst}\, M_\odot$ yr$^{-1}$ in the 0.3-3.0 keV temperature range from the analysis 
 of the MOS and pn data, respectively.  These values are dominated by the cold gas
 in the energy range 1.8-3.0 keV, while only upper limits can be obtained at temperatures below 1.8 keV.
The upper limits to the global mass deposition rate below 1.8 keV appear to be consistent with an SFR of 
$\sim 800 M_\odot$ yr$^{-1}$, as measured in the BCG by \citet{2013aMcDonald}, from the EPIC MOS data analysis within $1 \sigma$,
while the upper limits on $\dot M$ measured from EPIC pn data at temperatures below 1.8 keV are
significantly lower (more than $3\sigma$) than the SFR measured in \citet{2013aMcDonald}.

\item Considering the temperature range 0.3-1.8 keV, we found that  
MOS data analysis is consistent with $\dot M \sim 1000$ within $ 1\sigma$, while the pn data provide $\dot M < 400 M_\odot$ yr$^{-1}$ 
at $3\sigma$ c.l. 

\item Since the discrepancy between the MOS and pn data analyses cannot be
explained on the basis of currently known cross-calibration uncertainties between the two instruments nor 
of other sources of statistical noise, we argue that
additional calibration problems between EPIC instruments still need to be understood and properly treated.
The contribution of the cold gas with an average mass-deposition rate of $\sim 600 M_{\odot}$ yr$^{-1}$
in the EPIC data is about 5\% in the energy range 0.5-1.0 keV for our extraction region.  
This implies that any calibration uncertainty on the same order strongly affects the data.  
This conclusion is valid in the framework of the isobaric cooling model that we assumed here.  At present, we are unable to discuss whether the assumptions of different physical models for 
describing the cooling of the gas can change the picture and mitigate the difference between MOS and  pn analysis.

\item No line emission from ionization states below Fe XXIII is seen above $12 \AA $ in the RGS spectrum, and
the amount of gas cooling below $\sim 3$ keV has a formal best-fit value for the mass deposition rate of
$\dot M = 122_{-122}^{+343}$ $M_{\odot}$ yr$^{-1}$.  This result was confirmed by a direct comparison of 
the Phoenix RGS spectrum with A1835 in the overlapping spectral range.   This means that the mass deposition rate from RGS analysis is  
lower than the SFR of \citet{2013aMcDonald} in the BCG at the $2\sigma$ c.l. 

\item Current {\sl Chandra} data (from a short exposure of $\sim 10 $ ks) agree with our XMM-Newton analysis, 
but do not provide meaningful constraints on $\dot M$.  Deeper {\sl Chandra} data, thanks to the high angular resolution, are
expected to provide tighter constraints to the global mass deposition rate.

\item A careful analysis of the FIR SED based on    {\sl Herschel} data provided a value for the SFR in the BCG of  $530 \, M_\odot$ yr$^{-1}$
with an uncertainty of 15\%.  This revised estimate of the SFR  agrees very well with the $\dot M$ from EPIC MOS data and 
is consistent within $1 \sigma$ with $\dot M$ from the RGS analysis, while still inconsistent at 
more than $3 \sigma$ with the mass deposition rate found in EPIC pn data.  Our revised SFR therefore does not significantly change
the comparison between SFR and the global $\dot M$ in the Phoenix cluster.

\end{itemize}

To summarize, the range of $\dot M$ allowed by XMM-Newton data from EPIC MOS is consistent with the SFR observed
in the BCG, while EPIC pn and RGS data analyses suggest a mass deposition rate
of about $\sim SFR/3$, and the derived upper limit is inconsistent with the observed SFR at least at the $2\sigma$ level for RGS, and more than $3 \sigma$ for EPIC-pn. 
As a consequence, our results do not provide a final answer on the possible
agreement between the mass deposition rate of isobaric cooling gas in the core and the observed SFR in the BCG.  
As recently shown by  Molendi et al. (in preparation), $\dot M$ is often measured to be significantly lower
than the global SFR in central cluster regions for several strong cool-core clusters.  These findings
suggest that cooling flows may be  short-lived episodes, efficient in building
the cold mass reservoir that, on a different timescale and possibly with some delay, triggers
the star formation rate in the BCG.  To investigate whether this also occurs in the Phoenix cluster, or whether the Phoenix cluster actually hosts
the highest cooling flow observed so far with $\dot M \simeq SFR$, we must wait for a deep, high-resolution, spatially resolved X-ray spectral analysis 
to remove the stronger emission from the surrounding hot gas as best possible.

\begin{acknowledgements}
We acknowledge financial contribution from contract PRIN INAF 2012 ({\sl "A unique dataset
to address the most compelling open questions about X-ray galaxy clusters"}).
IB and JSS acknowledge funding from the European Union Seventh Framework Programme (FP7/2007-2013) under grant agreement no. 267251 "Astronomy 
Fellowships in Italy" (AstroFIt).  We thank Guido Risaliti for useful discussions on the XMM data reduction and analysis.  We also thank the anonymous referee for comments and suggestions
that significantly improved the paper.
\end{acknowledgements}

\bibliography{references_Clusters_PT}

\end{document}